\documentclass[pra,aps,groupedaddress,twocolumn,superscriptaddress,floatfix,showpacs]{revtex4}
\usepackage{epsfig}
\usepackage{mathrsfs}
\usepackage{amsmath}
\usepackage{textcmds}
\usepackage{amssymb}
\usepackage{mathptmx}
\allowdisplaybreaks[1]
\newcommand{\be}{\begin{equation}}
\newcommand{\ee}{\end{equation}}
\newcommand{\bea}{\begin{eqnarray}}
\newcommand{\eea}{\end{eqnarray}}
\newcommand{\ket}[1]{\left|#1\right\rangle}
\newcommand{\bra}[1]{\left\langle #1\right|}
\newcommand{\braket}[2]{\left\langle #1 | #2 \right\rangle}

\newcommand{\bc}{\begin{center}}
\newcommand{\ec}{\end{center}}

\renewcommand{\(}{\left(}
\renewcommand{\)}{\right)}
\renewcommand{\[}{\left[}
\renewcommand{\]}{\right]}
\newcommand{\forget}[1]{}

\newcommand{\re}{{\rm e}}
\newcommand{\rd}{{\rm d}}
\newcommand{\ri}{{\rm i\,}}
\begin{document}
\title{Arbitrary Coherent Superpositions of Quantized Vortices in Bose-Einstein Condensates \\
from Orbital Angular Momentum Beams of Light}
\author{Sulakshana Thanvanthri}
\affiliation{Hearne Institute for Theoretical Physics, Department of Physics \& Astronomy,
Louisiana State University,
Baton Rouge, Louisiana 70803-4001, USA}
\author{Kishore T. Kapale}
\email{KT-Kapale@wiu.edu}
\affiliation{Department of Physics, Western Illinois University, Macomb, Illinois, 61455-1367, USA}
\affiliation{Hearne Institute for Theoretical Physics, Department of Physics \& Astronomy,
Louisiana State University,
Baton Rouge, Louisiana 70803-4001, USA}
\author{Jonathan P. Dowling}
\affiliation{Hearne Institute for Theoretical Physics, Department of Physics \& Astronomy,
Louisiana State University,
Baton Rouge, Louisiana 70803-4001, USA}
\begin{abstract}
We recently proposed a scheme for the creation of coherent superpositions of vortex states in Bose-Einstein condensates (BEC) using  orbital angular momentum (OAM) states of light [Phys. Rev. Lett. {\bf 95}, 173601 (2005)]. Here we discuss further technical details of the proposal,  provide alternative, time-reversal-symmetric scheme for transfer of a superposition of OAM states of light to the BEC via a procedure analogous to the traditional STimulated Raman Adiabatic Passage (STIRAP) technique, and discuss an alternative trap configuration conducive for sustaining large charge vortices. Superpositions of OAM states of light, created using experimental techniques, can be  transfered to an initially nonrotating BEC via a specially devised Raman coupling scheme.  The techniques proposed here open up avenues to study coherent interaction of OAM states of light with matter. The study could also be employed for performing various quantum information processing tasks with OAM states of light\mdash including a memory for a quantum state of the initial superposition.
\end{abstract}
\maketitle

\section{Introduction}
The interaction of light, carrying quantized orbital angular momentum (OAM)~\cite{Padgett:2004}, with atomic or molecular matter is of considerable research interest~\cite{Muthikrishnan:2002,Babiker:2002,Alexandrescu:2006}. This quest is increasingly becoming feasible due to the progress that has been made in 
creation~\cite{Heckenberg:1992, Arlt:1998, Sueda:2004}, manipulation~\cite{Akamatsu:2003}, detection~\cite{Molina-Terriza:2002,Leach:2002},  and application~\cite{Grier:2003} of the orbital angular momentum (OAM) states of light. Nevertheless, due to the size-mismatch between the atoms and the spatial features of the phase structures in the OAM beams of light, it is difficult to couple the OAM degrees of freedom of light to the internal states of single atoms. As a result, some of the activity in this area is concentrated in the excitation of vortices in Bose-Einstein Condensates~\cite{Anderson:1995, Bradley:1995, Mewes:1996}.  Bose-Einstein Condensates (BEC)  are macroscopic coherent objects that are obtained by trapping atoms in particular internal states and cooling them further so that all the atoms have the same motional and internal states. Excitation of vortices in BECs have been traditionally achieved via stirring of the BEC cloud with a laser beam~\cite{Abo-Shaeer:2001}. These vortex states are fairly stable and could be candidates for qubits in Quantum Information~\cite{Nielsen:2000}, if appropriate means to manipulate them are developed. 

There exist several proposals to transfer pure OAM states of light to the BECs~\cite{Marzlin:1997, Nandi:2004}; there is also a proposal to use BECs for storage of the OAM state of light~\cite{Dutton:2004}. This transfer of OAM of light to BEC is possible due to the coherent nature of the BEC and because the size of the BEC cloud is about the same as the distance over which the intensity variations occur in the beam of light carrying OAM. We recently proposed a scheme for creation of superposition of two counter-rotating vortex states (with same absolute value of the charge) in BECs via coupling of the OAM degrees of freedom of incident beam to the center-of-mass motion of the atoms~\cite{Kapale:2005}. To illustrate, transfer of a pure OAM state of light to a BEC cloud is an experimental reality~\cite{Andersen:2006}; nevertheless the experimental system involves different  linear momentum for the initial and final states of the BEC as opposed to different internal states in the scheme discussed in Ref.~\cite{Kapale:2005} and in this article. Furthermore, the experiment involves transfer of only pure OAM states to the BEC as opposed to a OAM superposition as discussed here. It is important to note that coherent transfer of a superposition state is the first check necessary to test the suitability of these macroscopic objects for quantum information. The results of our study reported here demonstrate the possibility of using vortex states in BEC as a qubit.

The counter-rotating vortex superposition  generated via our scheme is analogous to the counter-rotating persistent currents in  superconducting circuits~\cite{Nakamura:1999,Friedman:2000,Wal:2000}. These counter-rotating persistent currents in the superconducting systems have already been recognized as candidates for qubits for quantum information processing. Thus the application of the study presented here to the area of quantum information is clear.  Further applications of our scheme could be as a quantum memory of such superposition states. One can envision such a memory device to be useful in the quantum communication networks using OAM states of light. (See Ref.~\cite{Spedalieri:2004QKD} and references therein.) OAM states of light offer a higher dimensional Hilbert space to obtain extra security and dense coding of quantum information. The memory application would deem it necessary to have a time-reversal symmetric procedure to write the superposition in OAM states of light to BEC so that the superposition can be extracted at a later stage. Here we discuss further technical details of our earlier proposal~\cite{Kapale:2005} and also provide a STIRAP-like time-reversal symmetric mechanism~\cite{Shore:1992} for transfer of special OAM states of light to BEC. Furthermore, we discuss alternative trapping potential, which can support toroid-shaped condensate cloud, in order to increase the stability of the generated vortex superposition.

The article is organized as follows. In Sec.~\ref{Sec:OptOAMSuperposition} we discuss generation of arbitrary superposition of two counter-rotating  OAM components with the same absolute value of charge. This section also discusses a scheme for generation of arbitrary superposition of two arbitrarily charged OAM states of light. Next, in Sec.~\ref{Sec:OAMCoupling}, we discuss two independent methods of transferring the OAM superposition from the light beam to the vortex states of BEC in complete detail. We also consider an alternative trap configuration, the so-called mexican-hat trap to offer an example of trap that can sustain large charge vortices. In fact, we provide a very general set of equations to study the transfer of OAM superposition from light to BEC for any kind of trapping potential.  In Sec.~\ref{Sec:Detection} we discuss the detection of the BEC vortex superposition and present our conclusions in Sec.~\ref{Sec:Conclusion}. Some of the mathematical details are presented in Appendices A and B.

\section{\label{Sec:OptOAMSuperposition}Superposition of optical vortices}

To recollect, the OAM states of light are different
in their phase characteristics from the conventional gaussian light beams. They have azimuthal phase structure characterized by the 
quantized orbital angular momentum carried by each photon in the beam. In the simplest case of  one unit of orbital angular momentum, i.e. $\hbar$,  the light beam has a cork-screw type phase front. Thus, the phase continuously varies from 0 to $2 \pi$ along the azimuthal coordinate and there is a jump from $2 \pi$ to 0 along some radial line. The location of the radial line along which the phase jump (or phase discontuinity) occurs is time-dependent and it rotates continuously with time. The sense of rotation governs the sign of the the orbital angular momentum. Furthermore, an OAM state with angular momentum $\ell \hbar$  has $|\ell|$ number of azimuthal phase jumps across a cut taken in the beam path.  It has to be noted that this is an instantaneous description of the beam phase across a cut taken in its path. The whole structure is time dependent  and rotates along the beam axis in a clockwise or counter-clockwise direction as the beam propagates. The sign of $\ell$ corresponds to the sense of rotations of the phase fronts around the beam axis. In principle, there is no upper limit to the angular momentum value that can be imparted to a light beam. Therefore, OAM states of light themselves are very good candidates for quantum information processing, mainly in the area of quantum cryptography,  as they offer an essentially infinite Hilbert space to work with. This could result in increasing the security of the quantum-key distribution protocols tremendously.

Monochromatic beams with azimuthal phase dependence of the type $\exp(\ri \ell \phi)$, of which 
Laguerre-Gaussian (LG) laser modes are an example, have well-defined angular momentum of $\ell \hbar$ per photon.
The normalized Laguerre Gaussian mode at the beam waist $(z=0)$ and beam size
$w_0$ at the waist is given in 
cylindrical coordinates ($\rho,\phi,z$) by
\begin{align}
\mbox{LG}_{p}^{ \ell }(\rho, \phi)& = \sqrt{\frac{2 p !}{\pi(| \ell |+p)!}} \frac{1}{w_0} 
\left(\frac{\sqrt{2}\rho}{w_0}\right)^{| \ell |} L^{| \ell |}_p\left(\frac{2 \rho^2}{w_0^2}\right)\nonumber \\
&\qquad\exp{(-\rho^2/w_0^2)}\exp{(\ri  \ell  \phi)}\,,
\end{align}
where $L^{ \ell }_p(\rho)$ are the associated Laguerre polynomials,
\begin{equation}
L^{| \ell |}_p(\rho)=\sum_{m=0}^p (-1)^m \frac{(| \ell |+p)!}{(p-m)!(| \ell |+m)!m!}\rho^m\,.
\end{equation}
Here, $p$ is the number of non-axial radial nodes, and the index $ \ell $, the winding number, describes the helical structure of the wave front and the number of times the phase jumps occur as one goes around the beam path in the azimuthal direction.  In general, any mode function
$\psi(\rho,\phi,z)$ can be expanded into LG modes as
\begin{equation}
\psi(\rho,\phi,z)=\sum_{ \ell =-\infty}^{\infty}\sum_{p=0}^{\infty} A_{ \ell p} \mbox{LG}^{ \ell }_p(\rho, \phi,z)
\end{equation}
For further discussion we consider only pure LG modes with a definite charge $\ell$ and $p=0$, we denote such a state of the
light field by $\ket{\ell}$ such that
$
\braket{\mathbf r}{\ell} = \mbox{LG}^{\ell}_0(\rho, \phi).
$
Thus it can be easily seen that the states $\ket{+\ell}$ and $\ket{-\ell}$ (with $\ell$ being a whole number)
differ only in the sense of the winding of the phase; either clockwise or counter-clockwise. 

Now we discuss the first step of the proposal: creation of an arbitrary  superposition of two OAM states of light of the kind
$(\alpha\ket{\ell}+\beta\ket{\ell^\prime})/\sqrt{\alpha^2+\beta^2}$. 
First we discuss generation of a special kind of superposition state:
$(\alpha\ket{\ell}+\beta\ket{-\ell})$, with $\alpha^2+\beta^2=1$. 
Here the two components of the state have opposite sense of rotation.
It is well known that creation of a very general multicomponent superposition of OAM states of the kind $\sum_{\ell} c_{\ell} \ket{\ell}$ 
is a fairly straightforward procedure by using holographs or phase plates. Furthermore, a sorter of these OAM
states has also been demonstrated~\cite{Leach:2002}, which distinguishes between different orbital angular momentum components.  Thus, by using a mixed OAM-state generator in conjunction with an OAM-sorter one can easily obtain a pure OAM state with arbitrary $\ket{\ell}$.  

\begin{figure}[ht]
\includegraphics[scale=0.5]{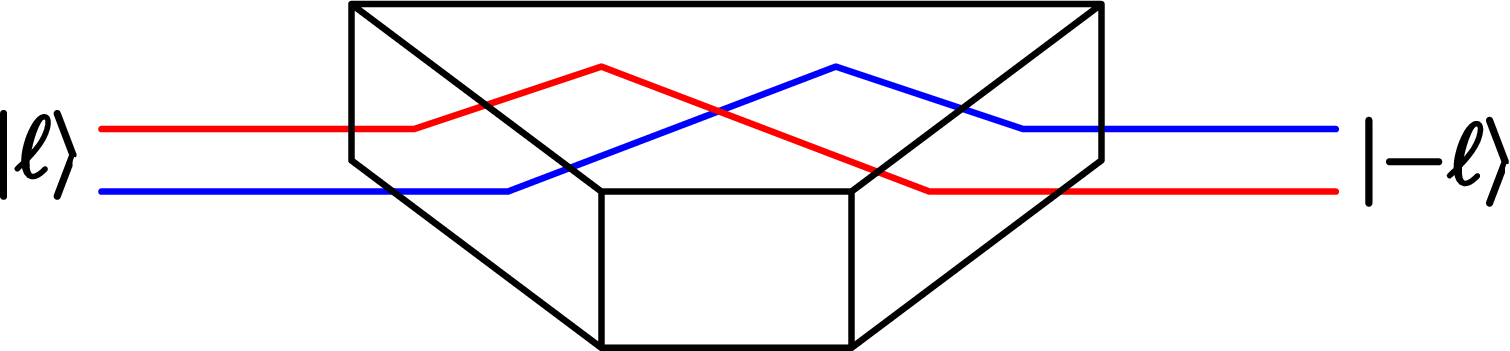}
\caption{\label{Fig:DovePrism} Dove prism as a sign shifter for the OAM states of light. The dove prism changes the handedness of the incident beam carrying OAM and causes a sign change in its winding number.}
\end{figure}

Next, we describe a technique that could be used to transform a $\ket{+\ell}$ state in to a $\ket{-\ell}$ state with
$\ell$ being a whole number. We illustrate this schematically in Fig.~\ref{Fig:DovePrism}. A common 
representation of the OAM states of light is through the Laguerre-Gaussian beams which have
circular cross-sections. We consider two rays (red and blue) as seen in Fig.~\ref{Fig:DovePrism} that lie
diametrically opposite to each other on the beam profile. As can be easily shown from the ray-diagram these rays
would interchange their places in the beam profile~\cite{Courtial:1997,Padgett:1999}. Extrapolating this to the whole beam it can be seen that 
passing a OAM state of light through a dove prism amounts to a anti-symmetrization of its azimuthal phase 
structure thus giving rise to a
$\ket{-\ell}$ state. It is also important to note, as pointed out by Padgett {\it et al.}~\cite{Padgett:1999}, that this rotation of the OAM state does not affect the polarization state of the light. To obtain complete conversion in the handedness of the input OAM, via the Dove prism, further care may be necessary as pointed out in Ref.~\cite{Gonzalez:2006}.

\begin{figure}[ht]
\includegraphics[scale=0.3]{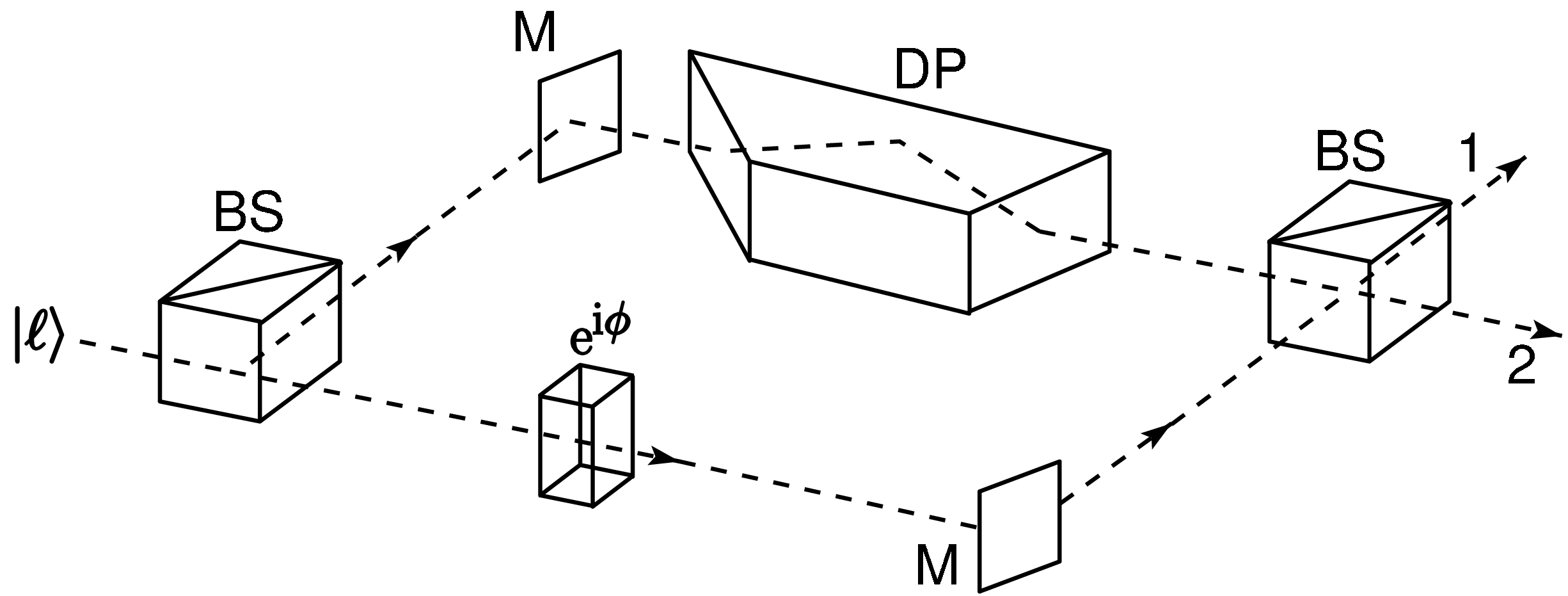}
\caption{\label{Fig:OAMS} Scheme for creation of superposition of the OAM states:
$(\ket{+\ell}+\ket{-\ell})/\sqrt{2}$.}
\end{figure}

Thus, starting with a $\ket{+\ell}$ state of light and passing it through a Mach-Zehnder type configuration shown in
Fig.~\ref{Fig:OAMS}, we obtain  $(\ket{\ell}+\ket{-\ell})/\sqrt{2}$ at one of the output ports of the
interferometer. By choosing the first beam splitter in the Mach-Zehnder interferometer as an $\alpha/\beta$ beam splitter and the second one is a 50/50 beam splitter one can generate a general two-component superposition $(\alpha \ket{\ell}+\beta \ket{-\ell})/\sqrt{|\alpha|^2 + |\beta|^2}$.

With the mathematical details discussed at length in Appendix~\ref{App:MachZehnder}, the process can be described as follows. The initial state of the light field before entering 
the first beam splitter is $\ket{\ell}$. After the passage through the first beam splitter, the beam is equally
split into two paths and the state is $\alpha \ket{\ell}_1 + \beta\ket{\ell}_2$. Beam 1 is the part of the beam that passes through the beam
splitter and beam 2 is the one that is reflected. Beam 2 later passes   through the dove prism
to undergo the transformation $\ket{\ell}\rightarrow\ket{-\ell}$ as discussed above. Thus, just before entering
the second beam splitter the state of the light field is $\alpha\ket{\ell}_1 + \beta\ket{-\ell}_2$, which becomes
$\alpha\ket{\ell} + \beta\ket{-\ell}$ after the recombination in the second beam splitter.
The complete Mach-Zehnder interferometer can be described via the transformation:
\begin{equation}
\(\begin{matrix} u_0 \ket{\ell} \\ 0 \end{matrix}\)\rightarrow\(\begin{matrix}  r^2 u_0 \ket{-\ell}  - t^2 {\rm e}^{\ri \phi} u_0\ket{\ell} \\   \ri t\,r \, u_0 \ket{-\ell}  + \ri \, r\,t  {\rm e}^{\ri \phi}\, u_0\ket{\ell} \end{matrix}\) \,.
\end{equation}
With the choice of 50-50 beam splitters we have $r=t=1/\sqrt{2}$ and $\phi=\pi$ to obtain the state at the output ports of the interferometer
\begin{equation}
\frac{1}{\sqrt{2}}\,u_0\(\begin{matrix} \ket{-\ell}  + \ket{\ell} \\   \ri(\ket{-\ell}  - \,\ket{\ell}) \end{matrix}\)
\end{equation}
Thus, by ignoring the port 2 and using the state from port 1 we obtain the required superposition state.

To generate arbitrary superposition of the kind $(\alpha \ket{\ell} + \beta \ket{\ell'})/\sqrt{\alpha^2 + \beta^2}$ scheme in Fig.~\ref{Fig:OAMS} needs to be changed slightly: (i) the input state needs to be  a normal Gaussian beam with no orbital angular momentum. (i) the dove prism needs to be replaced by a hologram to transfer the gaussian beam into the state $\ket{\ell'}$ and (iii) an extra hologram is added into the lower path to obtain conversion from the Gaussian beam into the state $\ket{\ell}$.

A superposition of two OAM states can be detected via a photo-detection scheme by taking a look at the spatial profile of the characteristic interference of the beams that are part of the superposition. For example, the interference pattern created by the superposition of $\ket{\ell}$ and $\ket{-\ell}$ is comprised of $2 \ell$ bright lobes that are equally distributed along the azimuthal direction.  The interference pattern of arbitrary superposition would contain $|\ell_1|+|\ell_2|$ lobes where $\ell_1$ and $\ell_2$ are the component OAM states. Thus the 
observation of the described interference pattern guarantees the creation of the coherent superposition of the appropriate states. Further details offered in Sec.~\ref{Sec:Detection} for detection of BEC vortex superposition can be applied here as well as the phase structures of the OAM superposition in light and vortex superposition in BEC are identical.

\section{\label{Sec:OAMCoupling}Coupling of optical vortex beams to BEC}
\begin{figure}[ht]
\centerline{\includegraphics[scale=0.35]{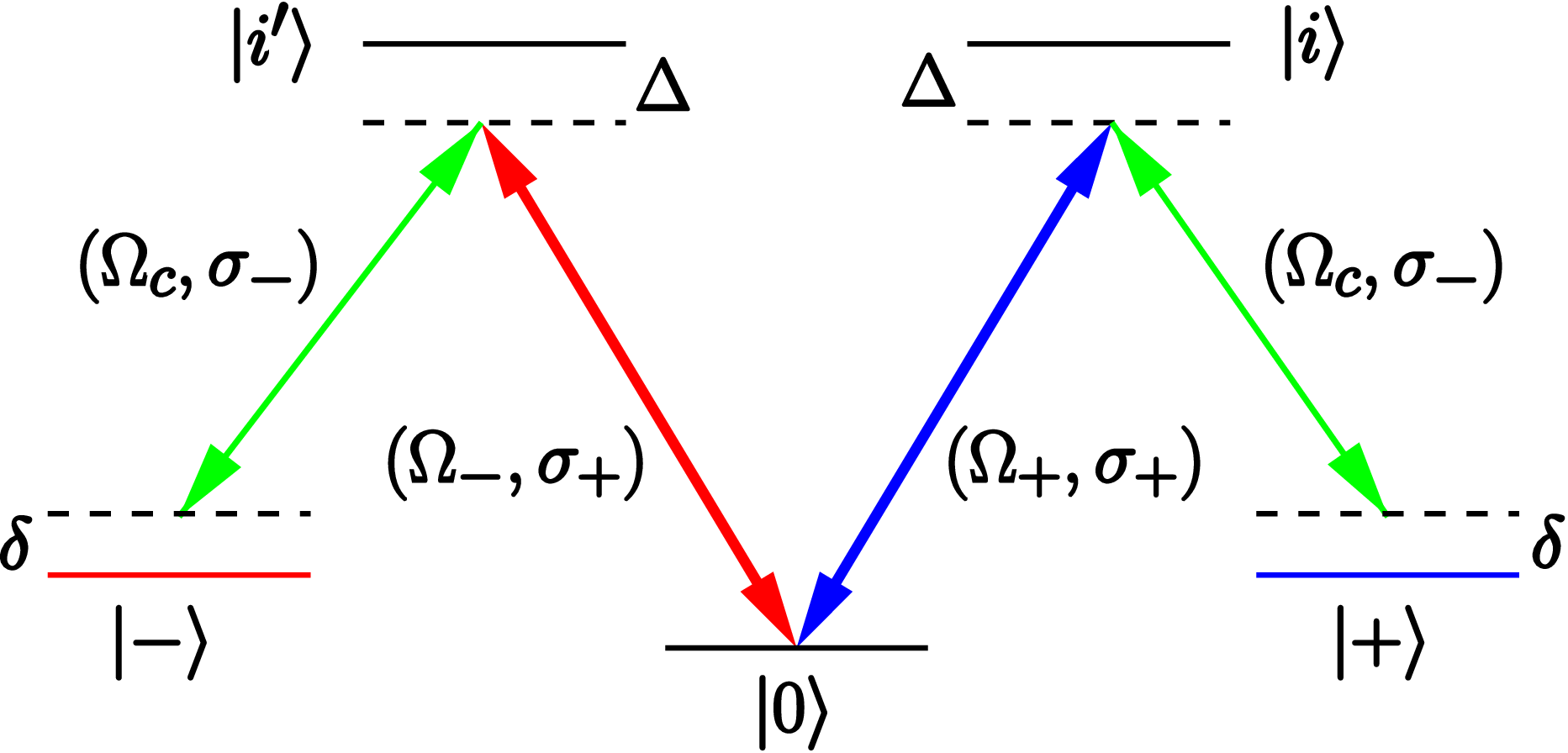}}
\caption{\label{Fig:LevelScheme}Internal level scheme of the trapped BEC atoms and the applied fields. The fields are specified through the corresponding Rabi frequencies and the polarization state. The fields are highly detuned from the excited states to avoid populating the excited states and ensure the stability of the condensate. The two-photon detunings $\delta$ are needed for one method of transfer of population from the non-rotating state to the vortex states but needs to be zero for the STIRAP like transfer mechanism.}
\end{figure}

In this section we discuss two methods to transfer  the generated  superposition of optical orbital angular momentum states to the states of atoms. Transfer of pure OAM states of light to BEC has been studied by Marzlin {\it et al.}~\cite{Marzlin:1997} and Nandi {\it et al.}~\cite{Nandi:2004}. Nevertheless, for the transfer of OAM superpositions one needs a special transfer scheme. The internal level scheme of the atoms and the transitions of interest are shown in the Fig.~\ref{Fig:LevelScheme}. The internal hyperfine quantum number of the initial non-rotating state $\ket{0}$ is $m_F=0$ and that of the final states $\ket{+}$ and $\ket{-}$ is $m_F=2$. Thus, the internal quantum number of both the final states is chosen to be the same to obtain a pure superposition of the vortex states while all rest of the quantum numbers needed to describe the two states are exactly identical. The intermediate states, $\ket{i}$ and $\ket{i'}$, through which Raman coupling between the initial and final states occurs, have the hyperfine quantum number $m_F=1$. The two components of the optical OAM states correspond to the Rabi frequencies $\Omega_+$ and $\Omega_-$ and both are $\sigma_+$ polarized. The coupling field is designated by the Rabi frequency $\Omega_c$ and it is $\sigma_-$ polarized. The OAM of the state $\ket{0}$ is zero and that of  $\ket{i}$ and $\ket{+}$ is $+\ell$ and that of $\ket{i'}$ and $\ket{-}$ is $-\ell$, where $\ell$ is a positive integer. Even though we are targeting generation of superposition of two counterrotating components, it can be quickly seen that this scheme can also be used as is for transfer of arbitrary superposition of two arbitrary OAM states of light to the BEC cloud.

Due to the various possibilities available for the internal state quantum numbers and the OAM quantum numbers there are five distinct states that the atoms in the BEC can have as a result of their interaction with the optical fields. Thus, the BEC cloud can be described through a five-component spinor $\{\Psi_0, \Psi_+, \Psi_-, \Psi_i, \Psi_{i'} \}$. All the internal states are assumed to be trapped by the trapping potential and thus there are mutual interactions among all the components. The optical fields couple various components of the spinor BEC, as shown in Fig.~\ref{Fig:LevelScheme}. Thus, we can write the evolution equations for the spinor components
\begin{align}
\ri\,  \dot{\Psi}_0  &=  \Omega_+^* \Psi_{i} + \Omega_-^*  \Psi_{i^\prime} + (\mathcal{H}_I/\hbar)\, \Psi_0 \nonumber \\
\ri  \dot{\Psi}_+  &=  \Omega_c^* \Psi_{i} + (\mathcal{H}_I/\hbar)\, \Psi_+ \nonumber \\
\ri\, \dot{\Psi}_-  &=  \Omega_c^* \Psi_{i^\prime}+ (\mathcal{H}_I/\hbar)\, \Psi_- \nonumber \\
\ri\, \dot{\Psi}_{i}  &=  - \Delta \Psi_i + \Omega_c \Psi_{+} +  \Omega_+ \Psi_{0} + (\mathcal{H}_I/\hbar)\, \Psi_{i} \nonumber \\
\ri\,  \dot{\Psi}_{i^\prime} &= - \Delta \Psi_{i^\prime} +  \Omega_c \Psi_{-} +  \Omega_- \Psi_{0} + (\mathcal{H}_I/\hbar)\, \Psi_{i^\prime}\,.
\label{Eq:Dynamics}
\end{align}
It is important to note the time derivatives appearing on the left-hand side of the equations. In general, the spinor components have both spatial and time dependence; nevertheless, for the optical evolution the time dependence is of prime importance. 
The optical coupling is governed by the Rabi frequencies $\Omega_c$ (for the coupling field) and $\Omega_{\pm}$ (for the OAM superposition). Here, $\Delta$ is the single-photon detuning. In setting up the above equations we have assumed that the two fields are in two photon resonance with the transitions $\ket{0}-\ket{+}$ and $\ket{0}-\ket{-}$, meaning the parameter $\delta$ appearing in Fig.~\ref{Fig:LevelScheme} is zero. This detuning can be added later on, if needed, as will become clear later.
Furthermore,
\begin{equation}
\label{HI}
\mathcal{H}_I  = (\mathcal{T} + \mathcal{V} -\mu ) + \eta ( |\Psi_0|^2 +  |\Psi_+|^2 +  |\Psi_-|^2 +|\Psi_{i}|^2 + |\Psi_{i^\prime}|^2 )
\end{equation}
where $\mathcal{T}$ is the Kinetic energy operator, $\mathcal{V}$ is the trapping potential of the BEC and $\mu$ is the chemical potential. The interatomic interaction strength is given by $\eta = 4 \pi \hbar ^{2} a N/m =  N U$, where $a$ is the s-wave scattering length, $N$ is the mean number of atoms in the cloud, and $m$ is the atomic mass of the atom. The absolute square terms in Eq.(\ref{HI}) thus correspond to the self-energy arising due to the interparticle interaction.

Noting the spatial profile of the LG beams, we obtain
\begin{equation}
\Omega_{\pm} ({\mathbf r}) = a_{\pm} \Omega_0\,  \re^{-r^2/w^2} \( \frac{\sqrt{2} r}{w}\)^{|\ell|} \re^{\pm\ri |\ell| \phi } \re^{\ri k z} 
\end{equation} 
with $\Omega_0$ being the Rabi frequency of  the atom-field interaction. $a_\pm$ are the probability amplitudes of the $\pm |\ell |$ vortices in the superposition state we want to create.
We assume that the Gaussian fall-off of the light intensity is over a length larger than the transverse size of the BEC cloud. As a result, we can ignore the gaussian term in the LG beam profile to arrive at
\begin{equation}
\Omega_{\pm} = a_{\pm}(\sqrt{2})^{|\ell |} \Omega_0\,\( \frac{x\pm \ri y}{w}\)^{|\ell|} \re^{\ri k z}\,.
\end{equation}
Note that we have incorporated the phase term inside the complex quantity $x\pm\ri y= r \cos\phi \pm \ri \sin \phi= r\re^{\pm \ri\phi}$. Similarly the gaussian fall-off of the intensity of the coupling beam (or $\Omega_c$)  along the transverse direction is ignored as the BEC cloud size is much smaller than the transverse beam profile.

Now,  we perform complete adiabatic elimination of the excited levels $\ket{i}$ and $\ket{i'}$. We are completely justified in using the adiabatic elimination procedure as: (i) the optical evolution happens on a time scale much faster than other processes in the BEC; (ii) the  optical fields are so detuned from the excited states that the excited states are never occupied.  Thus, by substituting $\dot{\Psi}_{i}=\dot{\Psi}_{i'}=0$ and eliminating $\Psi_{i}$ and $\Psi_{i'}$ from the Eq.~\eqref{Eq:Dynamics} we arrive at
\begin{widetext}
\begin{subequations}
\label{Eq:psidot}
\begin{align}
\label{Eq:psi0dot}
\ri \dot\Psi_{0} &= \frac{1}{\hbar}(\mathcal{T}+\mathcal{V}-\mu)\Psi_0 
                       +\frac{\eta}{\hbar} (  |\Psi_0|^2 +  |\Psi_+|^2 +  |\Psi_-|^2 ) \Psi_0 
		       + \frac{1}{\Delta}(|\Omega_+|^2 + |\Omega_-|^2)\Psi_0 
		       + \frac{1}{\Delta}\Omega_+^* \Omega_c \Psi_+ 
		       + \frac{1}{\Delta}\Omega_- ^*\Omega_c \Psi_- \\
\label{Eq:Psiplusdot}
\ri \dot\Psi_{+} &= \frac{1}{\hbar}(\mathcal{T}+\mathcal{V}-\mu)\Psi_+ 
                       +\frac{\eta}{\hbar} (\Psi_+|^2 +  |\Psi_-|^2 + |\Psi_0|^2  ) \Psi_+ 
		       + \frac{1}{\Delta}|\Omega_c|^2\Psi_+
		       +\frac{1}{\Delta}\Omega_+\Omega_c^* \Psi_0 \\
\label{Eq:Psiminusdot}
\ri \dot\Psi_{-} &= \frac{1}{\hbar}(\mathcal{T}+\mathcal{V}-\mu)\Psi_{-}   
                       + \frac{\eta}{\hbar} ( |\Psi_+|^2 + |\Psi_-|^2 
		       +|\Psi_0|^2 ) \Psi_- + \frac{1}{\Delta}|\Omega_c|^2\Psi_- 		      
		       + \frac{1}{\Delta}\Omega_-\Omega_c^* \Psi_0
\end{align}
\end{subequations}
\end{widetext}

Our ultimate interest lies in the temporal dynamics of the populations of various components of the BEC. Thus, the spatial part of the condensate is to be integrated out of the above equations. To deal with the spatial profiles of the spinor components of the BEC explicitly, we note that the general state of the BEC cloud can be written as
\begin{align}
\braket{\mathbf r}{\Psi} = \alpha(t) \braket{\mathbf r}{0} 
+ \beta(t) \braket{\mathbf r}{+} + \gamma(t) \braket{\mathbf r}{-}
\end{align}
where the spatio-temporal projections are given by
\begin{align}
\Psi_{0}({\mathbf r}, t) &= \alpha(t) \braket{\mathbf r}{0} = \alpha(t) \exp[\ri(\mu/\hbar - \kappa)t] \,\psi_{\rm g}({\mathbf r}) \nonumber \\
\Psi_{+}({\mathbf r}, t) &= \beta(t) \braket{\mathbf r}{+} = \beta(t)\exp[\ri(\delta+ \mu/\hbar - \kappa)t] \, \psi_{\rm v+}(\ell,{\mathbf r})\nonumber \\
\Psi_{-}({\mathbf r},t) &= \gamma(t) \braket{\mathbf r}{-} = \gamma(t)\exp[\ri(\delta+ \mu/\hbar - \kappa)t] \, \psi_{\rm v-}(-\ell,{\mathbf r})\,.
\label{Eq:Ansatz}
\end{align}
The two-photon detuning $\delta$ introduced in the phase factors is defined as $\delta = \nu_{\pm} - \nu_{c}  - \omega_{0\pm}$, where $\nu_{\pm}$ and $\nu_c$ are the angular frequencies of the optical field carrying OAM and the coupling field respectively, and  $\omega_{0\pm}$ is the energy level difference between the states 
$\ket{0}$ and $\ket{\pm}$ expressed as an angular frequency. 
Furthermore, the parameter $\kappa$, in general, depends on the interaction strength $\eta$ between the BEC atoms and the dimensional parameters of the BEC cloud, which in tern depend on the trapping potential. For example, for the Harmonic trapping potential $\kappa$ is given by
\begin{equation}
\kappa =\frac{\pi \hbar a_{\rm sc} N}{m (2 \pi)^{3/2} \hbar~ L_{\perp}^2 L_z} =\frac{\eta}{4 (2 \pi)^{3/2} \hbar~ L_{\perp}^2 L_z}\,.
\end{equation}
It can, however, be noted that as we are dealing with condensates consisting of single species, $\kappa$ need not be explicitely incorporated in the phase factors as it just causes a shift in the energy equally for all the component states of the BEC cloud and does not affect the population dynamics in a non-trivial manner.

The ansatz for the spatial profiles of the non-rotating ground state and the vortex states depend on the type of trapping potential we choose.  Before discussing different trapping potentials we take our general formalism a little further. After substituting the ansatz~\eqref{Eq:Ansatz} into the equations~\eqref{Eq:psidot} we obtain the temporal dynamics of the populations of the BEC components of interest:
\begin{widetext}
\begin{align}
\[ \ri\, \dot\alpha(t)-  (\frac{\mu}{\hbar}- \kappa)\, \alpha(t)\]\psi_{\rm g}(\mathbf{r})& = \frac{1}{\hbar}(\mathcal{T} +\mathcal{V} - \mu)\,\alpha(t)\, \psi_{\rm g}(\mathbf{r})   \nonumber \\
&+\frac{\eta}{\hbar} (|\alpha(t)|^2 |\psi_{\rm g}(\mathbf{r})|^2+ |\beta(t)|^2 |\psi_{\rm v+}(\ell,\mathbf{r})|^2 + |\gamma(t)|^2 |\psi_{\rm v-}(-\ell,\mathbf{r})|^2)\,\alpha(t)\, \psi_{\rm g}(\mathbf{r})  \nonumber \\
& + \frac{1}{\Delta}|\Omega_0|^2 \(\frac{\sqrt{2}r}{w}\)^{2\ell} \alpha(t)\, \psi_{\rm g}(\mathbf{r})  \nonumber \\
& + \frac{1}{\Delta} (a_+^* \re^{-\ri \ell \phi }  \,\beta(t) \,\psi_{\rm v+}(\ell,\mathbf{r}) +  a_-^*\re^{\ri \ell \phi }\, \gamma(t)\, \psi_{\rm v-}(-\ell,\mathbf{r})]\,\(\frac{\sqrt{2} r}{w}\)^{\ell}\Omega_0^*\, \Omega_c\,,
\nonumber \\
\[ \ri \, \dot\beta(t) -  (\frac{\mu}{\hbar}- \kappa+\delta)\, \beta(t)\] \psi_{\rm v+}(\ell,\mathbf{r})& = \frac{1}{\hbar}(\mathcal{T}+\mathcal{V}-\mu)\beta(t)\, \psi_{\rm v+}(\ell,\mathbf{r}) \nonumber \\
&+ \frac{\eta}{\hbar} \Bigl[|\alpha(t)|^2 |\psi_{\rm g}(\mathbf{r})|^2\,  + |\beta(t)|^2 |\psi_{\rm v+}(\ell,\mathbf{r})|^2+ |\gamma(t)|^2 |\psi_{\rm v-}(-\ell,\mathbf{r})|^2 \Bigr]\psi_{\rm v+}(\ell,\mathbf{r})\,\beta(t) \nonumber \\
&+ \frac{1}{\Delta} |\Omega_c|^2\,\beta(t)\, \psi_{\rm v+}(\ell,\mathbf{r})
+ \frac{1}{\Delta} a_+ \re^{\ri \ell \phi } \(\frac{\sqrt{2} r}{w}\)^{\ell} \Omega_0\, \Omega_c^*\, \alpha(t) \,\psi_{\rm g}(\mathbf{r})\,, 
\nonumber \\
\[ \ri \,\dot\gamma(t) -  (\frac{\mu}{\hbar}\,-\kappa+\delta)\, \gamma(t)\,  \]\psi_{\rm v-}(-\ell,\mathbf{r})&= \frac{1}{\hbar}(\mathcal{T}+\mathcal{V}-\mu)\gamma(t)\, \psi_{\rm v-}(-\ell,\mathbf{r}) \nonumber \\
&+ \frac{\eta}{\hbar} \Bigl[|\alpha(t)|^2 |\psi_{\rm g}(\mathbf{r})|^2+ |\beta(t)|^2 |\psi_{\rm v+}(\ell,\mathbf{r})|^2 + |\gamma(t)|^2 |\psi_{\rm v-}(-\ell,\mathbf{r})|^2\Bigr]\,\psi_{\rm v-}(-\ell,\mathbf{r})\, \gamma(t) \nonumber \\
&+ \frac{1}{\Delta} |\Omega_c|^2\,\gamma(t)\, \psi_{\rm v-}(-\ell,\mathbf{r}) + \frac{1}{\Delta} a_- \re^{-\ri \ell \phi } \(\frac{\sqrt{2} r}{w}\)^{\ell} \Omega_0\, \Omega_c^*\, \alpha(t) \,\psi_{\rm g}(\mathbf{r})\,.
\end{align}
\end{widetext}
Note that the exponential factors $\exp[\ri(\mu/\hbar - \kappa)t]$  and $\exp[\ri(\delta + \mu/\hbar - \kappa)t] $ have canceled as expected.  Also note that $\ell$ is taken to be a positive integer  and $\pm$ signs are used explicitly to easily distinguish the two counter-rotating vortex components; we will follow this convention from here onwards to avoid excessive use of the modulus sign. Now we formally perform the coordinate integrals to remove the spatial dependence.  In terms of the resulting integrals, which are summarized in Appendix B,  the rate equations are
\begin{widetext}
\begin{align}
\ri \dot{\alpha}(t) &=  (T_{g}+ V_{g}) \alpha(t) +(I_{gg} |\alpha(t)|^2 +I_{g+}(\ell)  |\beta(t)|^2+I_{g-}(\ell) |\gamma(t)|^2)\alpha(t)  
+ \frac{ |\Omega_0|^2}{\Delta}I_{gg}^{(2\ell)}(\ell)~ \alpha(t)
\nonumber \\
&\quad+ \frac{ \Omega_0^*\, \Omega_c}{\Delta}\(I_{g+}^{(\ell)} (\ell)~ a_{+}^* \beta(t) + I_{g-}^{(\ell)} (\ell)~ a_{-}^* \gamma(t)\)\,,
\nonumber \\
\ri \dot{\beta}(t) &= [T_{+}+V_{+}(\ell)+\delta] \beta(t) + (I_{g+}(\ell)~|\alpha(t)|^{2}  +I_{++}(\ell)~|\beta(t)|^2 +I_{+-}(\ell)~|\gamma(t)|^2  )\beta(t)+ 
 \frac{ |\Omega_c|^2}{\Delta} \beta(t)
+ \frac{\Omega_0\, \Omega_c^*}{\Delta} I_{+g}^{(\ell)}(\ell)~ a_{+}\, \,\alpha(t)\,,
 \nonumber \\
\ri \dot{\gamma}(t) &=[T_{-}+V_{-}(\ell)+\delta] \gamma(t) + 
(I_{g-}(\ell) |\alpha(t)|^{2} +I_{+-}(\ell) |\beta(t)|^2 +I_{--}|\gamma(t)|^2 )\gamma(t)+  \frac{ |\Omega_c|^2}{\Delta}\gamma(t)
+ \frac{\Omega_0\, \Omega_c^*}{\Delta} I_{-g}^{(\ell)}(\ell)~ a_{-}\,\alpha(t)\,.
\label{Eq:RateEqBasicFormal}
\end{align}
\end{widetext}
Note that the spatial integrals are denoted by the letter $I$ with different subscripts, superscripts and in most cases with the argument of $\ell$ to properly distinguish them. These integrals are explicitly written out in the Appendix B.  

To note, the above set of equations is very general and applicable to wide variety of trapping potentials, provided the integrals are evaluated by appropriately taking the spatial forms of the wavefunctions, $\psi_g(\mathbf{r}), \psi_{v\pm}(\mathbf{r})$, suitable for the trapping potential. Furthermore the equations can also be applied to situations where the applied OAM and coupling fields have a certain time profile. 
In case the spatial ansatz for the BEC wavefunction in the Thomas-Fermi limit then the kinetic energy terms can be ignored as we do in the case of the mexican hat potential. The application of these equations shall become clear as we discuss various cases below.

In the forthcoming subsection we look at two different trapping potentials and two different transfer mechanisms of OAM to a BEC.

\subsection{Harmonic Potential Trap}

The harmonic potential we consider is of the form,
\begin{equation}
V(x, y, z) = \frac{1}{2} m [\omega_{\perp}^{2} (x^{2}+y^{2}) + \omega_{z} z^{2}]\,.
\end{equation}
We assume a pancake shaped BEC cloud for which $\omega_{\perp} <\omega_{z}$ and as a result, $L_{\perp} > L_z$.
The spatial wavefunctions for the BEC trapped in this kind of trap can be taken to be,
\begin{align}
\psi_g({\mathbf r})&= \exp{\{-(1/2)[(r/L_{\perp})^2 + (z/L_{z})^2 ] \}}/(\pi^{3/4} L_{\perp} L_{z}^{1/2})\nonumber \\
 \psi_{v\pm}(\pm\ell,{\mathbf r})&=(x\pm \ri y)^{|\ell|} \psi_g({\mathbf r})/ (\sqrt{|\ell |!} \, L_{\perp}^{|\ell |} )\nonumber \\ &= r^{|\ell|} e^{\pm \ri \ell \phi} \psi_g({\mathbf r})/ (\sqrt{|\ell |!} \, L_{\perp}^{|\ell |} )\ .
\end{align}
The treatment, so far, is very general and would work for any value 
$\ell$ of the OAM of the incident light. To understand the dynamics more clearly we restrict ourselves to a particular value of the OAM, $\ell=2$. However, one may note that even though the spatial integrals would have different values for different $\ell$ values, the general idea remains the same. 

We start with Eq.~\eqref{Eq:RateEqBasicFormal} and substitute the spatial integrals for the particular case of the 3D harmonic trap, which are listed in Appendix B, to arrive at:
\begin{widetext}
\begin{align}
\ri\dot{\alpha}(t)&= -\kappa\, \alpha(t) + (\frac{1}{4}\omega_z + \frac{1}{2} \omega_{\perp} ) \alpha(t) + \frac{1}{2} \omega_{\perp} \alpha(t) + 3 \kappa |\alpha(t)|^2  \alpha(t) + \kappa\(|\alpha(t)|^2 + |\beta(t)|^2 + |\gamma(t)|^2 \) \alpha(t) \nonumber \\
&+  \frac{8}{\Delta}|\Omega_0|^2 
 \(\frac{\hbar}{m \omega_{\perp} w^2}\)^2 \alpha(t) 
+\frac{2 \sqrt{2}}{\Delta}\Omega_0^* \Omega_c  \frac{\hbar }{m \omega_{\perp}w^2}( a_{+}^* \beta(t) + a_{-}^*\gamma(t))\,, \nonumber \\
\ri \dot{\beta}(t)&= (-\delta - \kappa)\beta(t) + (\frac{1}{4}\omega_z + \frac{3}{2} \omega_{\perp}  ) \beta(t) + \frac{3}{2} \omega_{\perp} \beta(t) +  \kappa (|\alpha(t)|^2  + |\beta(t)|^2 +|\gamma(t)|^2 ) \beta(t)  
+  \frac{1}{2}\kappa ( |\beta(t)|^2 + |\gamma(t)|^2 )\beta(t) \nonumber \\
&+ \frac{1}{\Delta} |\Omega_c|^2 \beta(t) 
+ \frac{2\sqrt{2}}{\Delta} \Omega_{0}\Omega_{c}^* a_{+} \frac{\hbar}{m \omega_{\perp}w^2}\alpha(t)\,,\nonumber \\
\ri \dot{\gamma}(t)&= (-\delta - \kappa)\gamma(t) + (\frac{1}{4}\omega_z + \frac{3}{2} \omega_{\perp}  ) \gamma(t) + \frac{3}{2} \omega_{\perp} \gamma(t) +  \kappa (|\alpha(t)|^2  + |\beta(t)|^2 +|\gamma(t)|^2 ) \gamma(t)  +  \frac{1}{2}\kappa ( |\beta(t)|^2 + |\gamma(t)|^2 )\gamma(t) 
\nonumber \\
&+ \frac{1}{\Delta} |\Omega_c|^2 \gamma(t) 
+ \frac{2\sqrt{2}}{\Delta} \Omega_{0}\Omega_{c}^* a_{-} \frac{\hbar}{m \omega_{\perp}w^2}\alpha(t)\,.
\label{Eq:RateEqBasic}
\end{align}
\end{widetext}
Note the appearance of the Harmonic oscillator trap parameters.
At this stage, we note that, $|\alpha(t)|^2  + |\beta(t)|^2 +|\gamma(t)|^2=1$,  and eliminate terms that are common in all the equations as they do not give any non-trivial contributions to the population dynamics of the system. We also add an extra assumption to relate the properties of the BEC trap and the profile of the OAM carrying light beam such that ${2 \sqrt{2} \hbar}/({m \omega_{\perp} w^2})=1$. Note that this only scales the optical coupling constant appearing in the equations to a manageable number and does not change the physics in any way. Thus, we arrive at a general set of equations coupling a non-rotating state of BEC to two counter-rotating vortex states carrying charges $2$ and $-2$ respectively:
\begin{widetext}
\begin{align}
\ri\dot{\alpha}(t)&=  3 \kappa |\alpha(t)|^2  \alpha(t) +  \frac{1}{\Delta}|\Omega_0|^2 \alpha(t) 
+\frac{1}{\Delta}\Omega_0^* \Omega_c ( a_{+}^* \beta(t) + a_{-}^*\gamma(t))\,, \nonumber \\
\ri \dot{\beta}(t)&= (\delta+ {2} \omega_{\perp}) \beta(t) 
+  \frac{1}{2}\kappa ( |\beta(t)|^2 + |\gamma(t)|^2 )\beta(t) + \frac{1}{\Delta} |\Omega_c|^2 \beta(t) 
+ \frac{1}{\Delta} \Omega_{0}\Omega_{c}^* a_{+}\alpha(t)\,,\nonumber \\
\ri \dot{\gamma}(t)&= (\delta + {2} \omega_{\perp}) \gamma(t)  
+  \frac{1}{2}\kappa ( |\beta(t)|^2 + |\gamma(t)|^2 )\gamma(t) 
+ \frac{1}{\Delta} |\Omega_c|^2 \gamma(t) 
+ \frac{1}{\Delta} \Omega_{0}\Omega_{c}^* a_{-} \alpha(t)\,.
\label{Eq:RateEqBasicHT}
\end{align}
\end{widetext}
Starting with these equations, we develop two methods for transfer of optical OAM to a BEC. It needs to be noted that the dynamics, as seen in the above equations, cannot have a steady state due to large detunings from the excited states and exclusion of radiative decays. This regime of parameters is, of course, necessary to make sure that the BEC atoms do not populate the excited state, where the BEC may not survive. Thus, to obtain the complete population transfer in a robust manner we need to construct dynamics that would not cause the population to oscillate between the initial state and the final states but would rather show a one-way trend in the direction the population is moving. This, of course, could be done in both a time-reversal symmetric or non-symmetric manner. We study both these avenues.  The non time-reversal symmetric scheme involves linear chirp of the coupling field, meaning a time dependent two-photon detuning $\delta$. The time-reversal symmetric scheme is based on the traditional STIRAP population transfer mechanism. To note, the final state is not a single state but a superpositon of two states. We consider these two techniques one by one in the following subsections.

\subsubsection{Superposition of Vortices in BEC via frequency chirp}

To study the generation of the superposition of two counter-rotating vortices in BEC we start with Eq.~\eqref{Eq:RateEqBasicHT} and introduce two simplifications given by $\Omega_0=\Omega_c$ and $|\Omega_0|^2/\Delta = \omega_{\perp}$. There is no loss of generality in these assumptions and appearance of any other constant that related the light beam intensity parameter with the transverse trap frequency would demonstrate the same physics except the time scales may be a little different. Our aim here is, of course, to demonstrate that superposition of vortex states can be created in a robust manner despite the presence of inter-particle interactions as normally observed in BEC. The same formalism can be applied to the experimental situations to rigorously determine the time scales over which the population transfer occurs. In any case, the optical time evolution time scales ($\mu$s) are much shorter than the time scales for spatial evolution of the BEC (ms).

\begin{figure}[ht]
\includegraphics[scale=1.0]{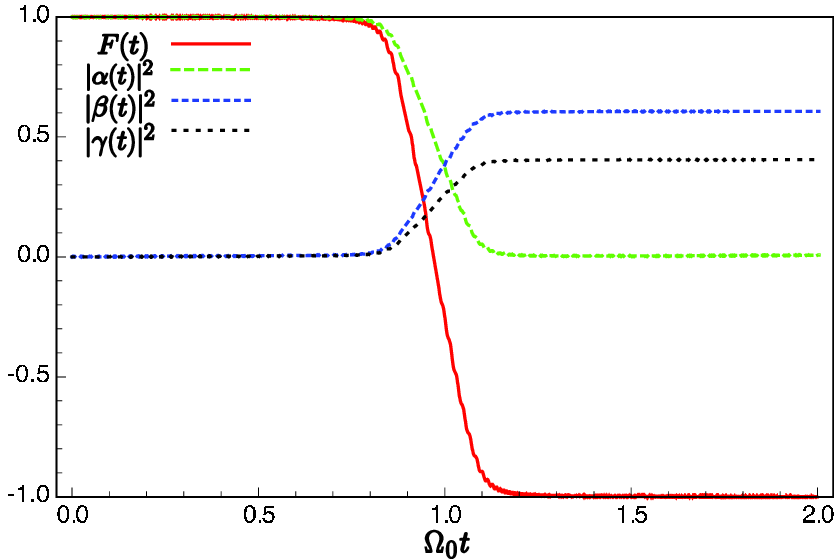}
\caption{\label{Fig:PlotChirp} Generation of the vortex state superposition via the linear frequency chirp technique. An equal superposition of the component vortex states is generated here starting with a non-rotating state of the BEC. The time plotted on the x-axis is measured in seconds. The linear chirp used for complete population transfer is given by the two-photon detuning $\delta(t) = C (1-\Omega_{0} t)$, where $C=2 \Omega_0$ and  $\Omega_{0} = 3$kHz. Other parameters are $\omega_{\perp}=132$ Hz, $a = 5$ nm, $L_{\perp}=2.35 \mu$m, $L_z=1.4\mu$m and $\kappa=1.7$kHz.}
\end{figure}

We obtain a simple set of equations governing the population of the three components of the BEC:
\begin{eqnarray}
\ri \dot{\alpha}(t) &=&  3 \kappa |\alpha(t)|^2 \alpha(t) + \omega_{\perp} 
\(a_{+}^* \beta(t) + a_{-}^* \gamma(t)\)\,
\nonumber \\
\ri \dot{\beta}(t) &=& (\delta  + 2 \omega_{\perp})\,\beta(t) + 
\frac{1}{2} \kappa  \(|\beta(t)|^2 +|\gamma(t)|^2 \) \beta(t)
\nonumber \\
& &+ \omega_{\perp}\, a_{+}\, \alpha(t)\,,
 \nonumber \\
\ri \dot{\gamma}(t) &=&\(\delta  + 2 \omega_{\perp} \)\gamma(t) + 
\frac{1}{2} \kappa  \(|\beta(t)|^2 +|\gamma(t)|^2 \) \gamma(t)
\nonumber \\ & &+ \omega_{\perp}\, a_{-}\, \alpha(t)\,.
\end{eqnarray}
The optical vortex superposition can be transferred to the BEC via a continuously chirped control pulse $\Omega_{c}$. The chirp is modeled by introducing linear time dependence in the two-photon detuning in the form $\delta(t)= C(1-\Omega_0 t)$, where $C$ is some appropriate constant. We discuss the solution of the above equation in Fig.~\ref{Fig:PlotChirp}. 

The solid line in the plot is the transfer function 
\begin{equation}
F(t) = |\alpha(t)|^2 - |\beta(t)|^2 - |\gamma(t)|^2
\label{Eq:Ft}
\end{equation}
which varies from the value 1, when all atoms are in state $\ket{0}$, to the value $-1$, when all the atoms are in state $(\sqrt{3}\ket{+2}+\sqrt{2}\ket{-2})/\sqrt{5}$, which is the required vortex superposition for 60:40 division of the population among the two vortex components. The population of state $\ket{0}$ is shown by a dashed line and it varies from 1 to 0; whereas the populations of the states $\ket{+2}$ and $\ket{-2}$  denoted by $|\beta(t)|^2$ and $|\gamma(t)|^2$ are given by dotted line which vary from 0 to 0.6 and 0.4 respectively, as expected. The coupling field frequency is to be varied linearly such that it sweeps from one side of the two-photon resonance to the other side to facilitate complete population transfer. For modeling purposes we vary the two-photon detuning $\delta$ so that the system sweeps through the two-photon resonance. 

 An alternative method we discuss below, to accomplish the OAM transfer, is similar to the traditional STIRAP technique of the counter-intuitive pulse sequence. This technique has been applied for the transfer of a pure OAM state to the BEC.

\subsubsection{BEC vortex superposition via STIRAP-like pulse sequence}
\label{sec:STIRAP}

The only change needed here from the previous subsection is to make the Rabi frequencies time dependent and to deploy counter-intuitive pulse sequence that is required to obtain the population transfer. 


We add the time dependence to the Rabi frequencies by via
\begin{equation}
\Omega_0 = |\Omega_0| f(t) \quad \mbox{and}\quad \Omega_c = |\Omega_0| g(t)\,,
\end{equation}
where the temporal profiles of the beams are taken to be of the form:
\begin{align}
f(t) = f_0\, \re^{-\left(\frac{t-t_1}{\sigma_1}\right)^2}\,, \quad g(t) = g_0\, \re^{-\left(\frac{t-t_2}{\sigma_2}\right)^2}\,.
\label{Eq:TimeProfile}
\end{align}
With these modifications and that $\delta=0$ the rate equations~\eqref{Eq:RateEqBasicHT} become

\begin{widetext}
\begin{align}
\ri \dot{\alpha}(t) &= \frac{1}{\Delta} |\Omega_0|^2 f(t)^2 \alpha(t) + 3 \kappa |\alpha(t)|^2 \alpha(t) + \frac{1}{\Delta} |\Omega_0|^2\,f(t)\,g(t)\, \(a_{+}^* \beta(t) + a_{-}^* \gamma(t)\)\,,
\nonumber \\
\ri \dot{\beta}(t) &= 2 \omega_{\perp}\,\beta(t) + 
\frac{1}{2} \kappa  \(|\beta(t)|^2 +|\gamma(t)|^2 \) \beta(t)
+ \frac{1}{\Delta} |\Omega_0|^2 g(t)^2 \,\beta(t)
+ \frac{1}{\Delta} |\Omega_0|^2\,f(t)\,g(t)\, a_{+}\, \,\alpha(t)\,,
\nonumber \\
\ri \dot{\gamma}(t) &=2 \omega_{\perp}\,\gamma(t) + \frac{1}{\Delta} |\Omega_0|^2 g(t)^2 \,\gamma(t)+ 
\frac{1}{2} \kappa  \(|\beta(t)|^2 +|\gamma(t)|^2 \) \gamma(t)
+ \frac{1}{\Delta} |\Omega_0|^2\,f(t)\,g(t)\, a_{-}\,\alpha(t)\,.
\label{Eq:STIRAP}
\end{align}
\end{widetext}

\begin{figure}[ht*]
\includegraphics[scale=1.0]{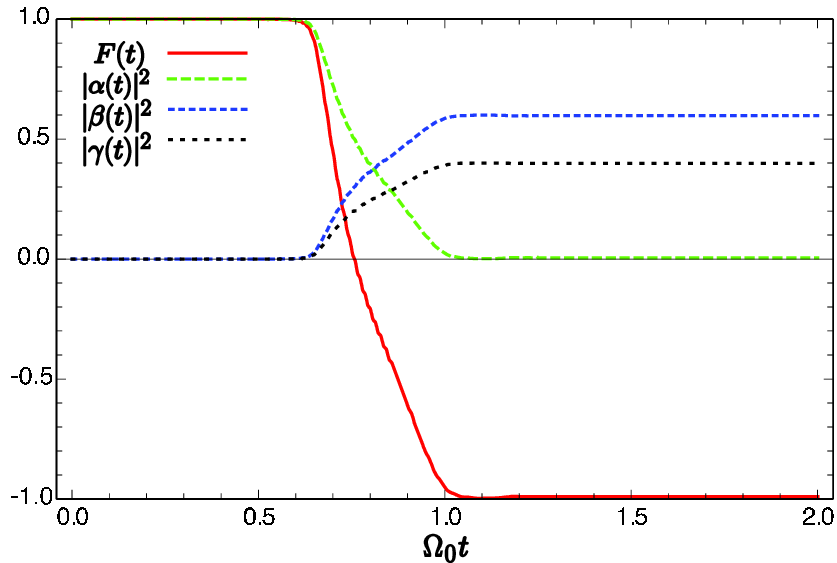}\hskip1cm
\caption{\label{Fig:PlotSTIRAP} Generation of the vortex state superposition: Results of the numerical solutions of the equations for STIRAP scheme shows the superposition 60:40 of the $\ket{+}$ and $\ket{-}$ vortex states. 
 The pulse profiles $f(t)$ and $g(t)$ are of the form in Eq.~\eqref{Eq:TimeProfile} with $\sigma_1=\sigma_2=0.25$ and $t_1=1.0$ and $t_2=0.5$ in the units of $1/\Omega_0$ with $f_0/g_0 = 0.5$ to obtain the complete transfer of the non-rotating BEC into the vortex superposition. (See Eq.~\eqref{Eq:STIRAP}.) Other parameters are $\Omega_0=2\times10^5$Hz, $\Delta=10 \Omega_0$ and $\delta=0$. The quantity on the $x$ axis is the scaled time $\Omega_0 t$.}
\end{figure}

These equations can be solved in a straightforward manner. The typical results are summarized in Fig.~\ref{Fig:PlotSTIRAP}. It can be seen that the transfer function plot signifies a complete transfer from initial state $\ket{0}$ to a 60:40 superposition of the vortex states $\ket{+2}$ and $\ket{-2}$. Any combination of superpositions (50/50, 60/40 or 80/20) can be generated via the counter intuitive pulse sequence---the coupling-field pulse (time profile $g(t)$ and pulse center $t_2$) comes before the OAM superposition pulse (time profile $f(t)$ and pulse center $t_1$).

\begin{figure}[ht]
\includegraphics[scale=0.7]{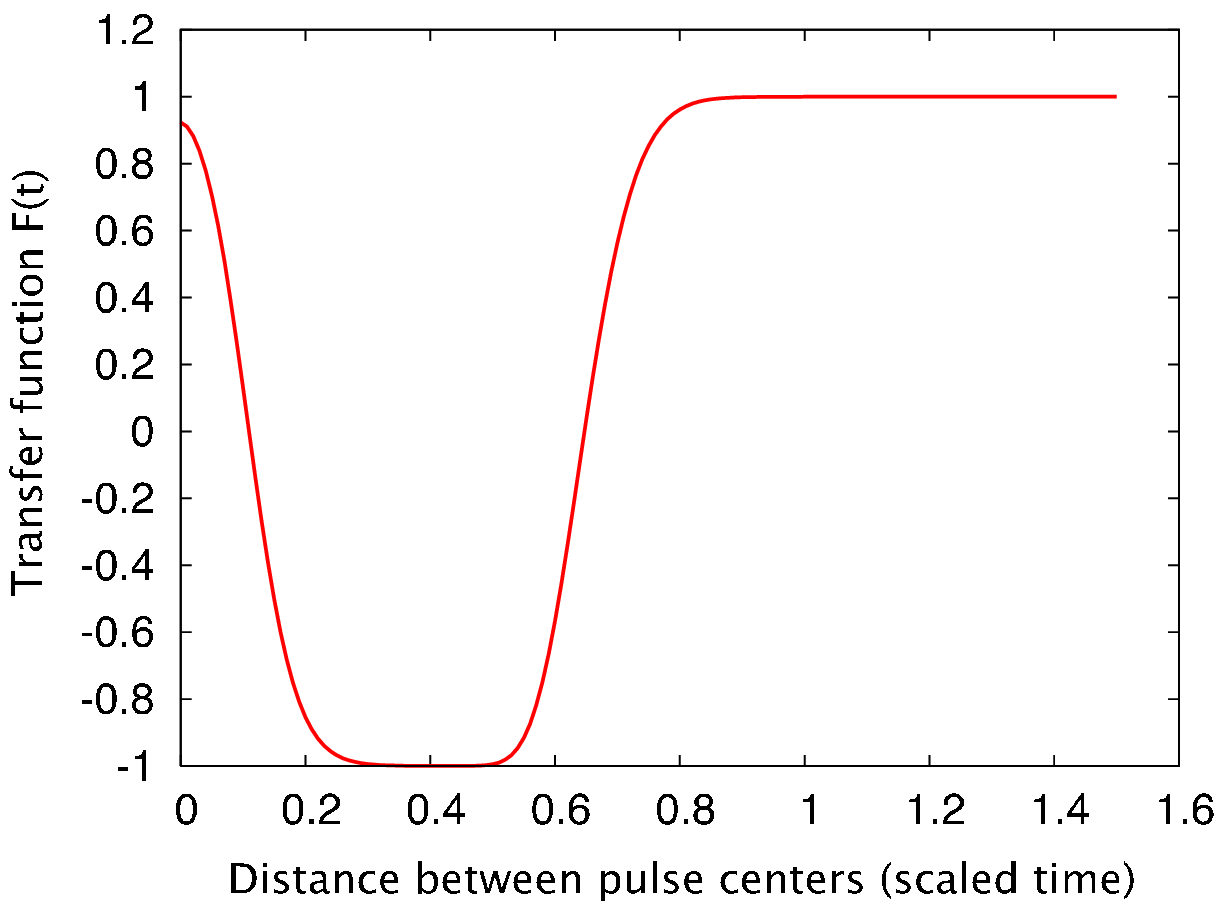}
\caption{\label{Fig:TransferFunctionVsOverlap} Plot of the transfer function (Eq.~\eqref{Eq:Ft}) versus the distance between the pulse centers  $t_1-t_2$. Note that $t_1-t_2>0$ (or $t_1>t_2$) suggests the counter-intuitive pulse sequence needed for the STIRAP based population transfer process. The other parameters are identical to those noted in the caption of the Fig.~\ref{Fig:PlotSTIRAP}}
\end{figure}

To understand the pulse overlap necessary to obtain complete population transfer from the initial non-rotating state to the final vortex superposition state via the STIRAP process, we plot the transfer function $F(t)$ vs the distance between the pulse centers $t_1-t_2$ in Fig. \ref{Fig:TransferFunctionVsOverlap}.
The value of $F(t)$ close to 1 means the population is primarily in the non-rotating ground state and the transfer is not efficient. Whereas, $F(t)=-1$ means complete population transfer to the vortex superposition has occurred. The pulse parameters are chosen to be exactly identical; therefore, the distance between the pulse center could be taken as the measure of the pulse overlap. For the light pulses arriving at about the same time the transfer is inefficient. However, the population transfer improves as the pulse separation increases a little. One obtains complete transfer for the range $\{0.3, 0.5\}$ for the separation between the pulses; beyond that the pulses are so far away from each other that the STIRAP process does not work. 

We would further like to point out that our numerical study suggests that within the parameter range used to obtain Figs.~\ref{Fig:PlotChirp} and ~\ref{Fig:PlotSTIRAP} the excited states populations are negligible and we are justified in using the adiabatic elimination of the excited states.

\subsection{Mexican hat trapping potential}

\begin{figure}[ht]
\includegraphics[width=0.9\columnwidth]{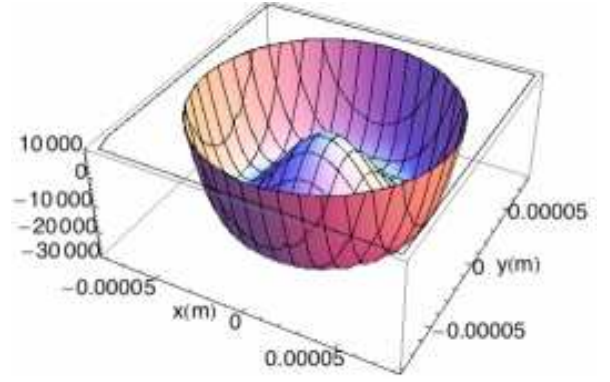}
\caption{\label{Fig:Pot}  The mexican hat potential in two dimensions with $\sigma = 2.0$ and $\lambda = 0.005$. }
\end{figure}

\begin{figure}[ht]
\includegraphics[scale=0.7]{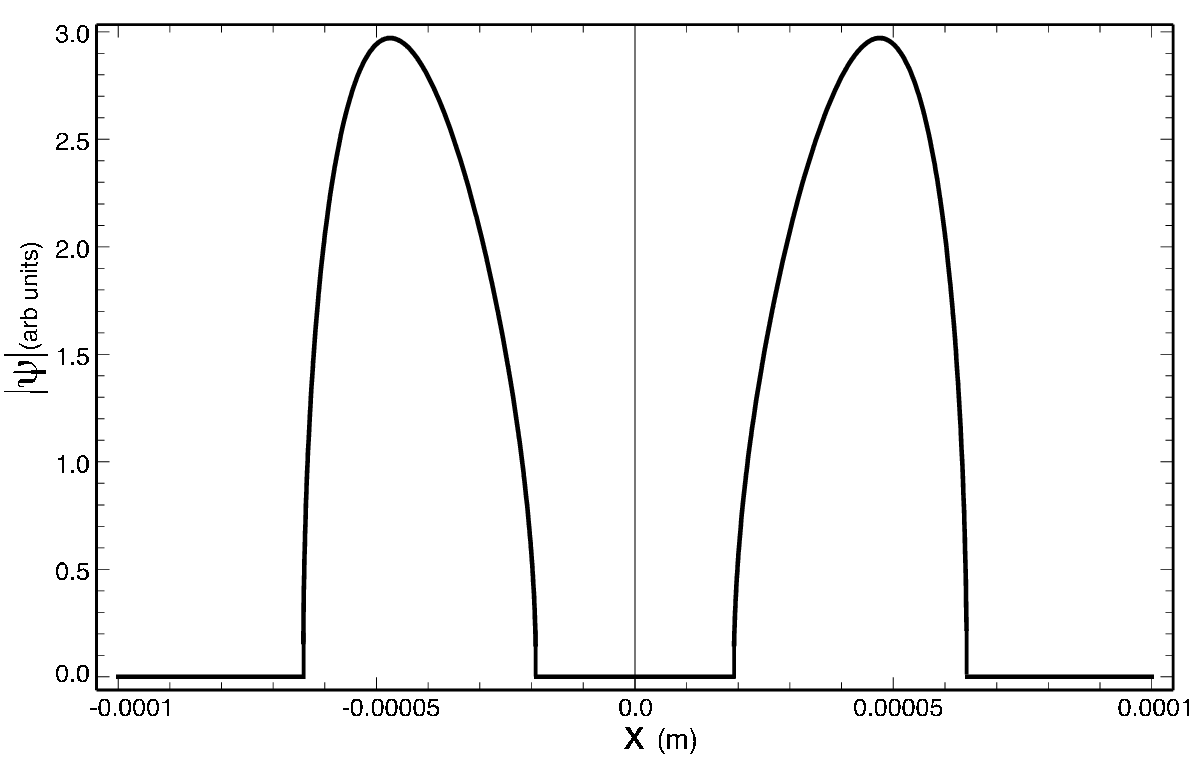}
\caption{\label{Fig:WF} The Thomas-Fermi spatial wavefunction of the BEC in a Mexican Hat potential in one dimension. }
\end{figure}

We now study the transfer of OAM to a BEC cloud when the trapping potential is shaped like a Mexican hat (Sombrero) in the $x-y$ plane (See Fig.~\ref{Fig:Pot}). The advantage of this potential is the toroidal symmetry it offers. The toroidal trap configurations naturally support toroid shaped BEC cloud and as a result can sustain large vortices without disintegration into several single charge vortices. This offers stability to the vortex superpositions we are aiming to generate. The trapping potential, in this case, is of the form \cite{Stringari:2006}

\begin{equation}
V(\rho,z)  =  -\frac{1}{2} \sigma m \omega_{\perp}^{2} \rho^{2} + \frac{1}{4} \lambda \(\frac{m^{2} \omega_{\perp}^{3}}{\hbar}\) \rho^{4} + \frac{1}{2} m \omega_{z}^{2} z^{2}
\end{equation}
where $\sigma$ and $\lambda$ are dimensionless parameters. The potential is harmonic in the z-direction. The dynamics of the OAM transfer to BEC can be studied in the Thomas-Fermi (TF) approximation where the kinetic energy of the BEC cloud is neglected. In this approximation, the spatial part of the BEC wavefunction is given by
\begin{align}
\psi(\ell,\rho,\phi, z)& =  \(\frac{1}{L_{\perp}\sqrt{L_{z}}}\)  \(\frac{1}{\sqrt{|\ell|!}}\) \(\frac{\rho}{L_{\perp}}\)^{|\ell|}\re^{-\frac{z^{2}}{2 L_{z}^{2}}}\nonumber \\
& \mbox{Max}\[ \mbox{\bf Re}\(\sqrt{\frac{\mu-V(\rho,0)}{\eta}}\),0 \] \re^{\ri \ell \phi} \,.
\end{align}
Here we have assumed that the wavefunction in $z$-direction has a Gaussian form and has little effect on the vortex dynamics  of interest in the transverse direction~\cite{Ketterle:2001}. The spatial profile of the above wavefunction is shaped like a toroid (or a donut) with a hole in the center. Thus, the two radii (inner and outer) are necessary to describe the shape of the BEC cloud.
The radii can be found by setting $|\psi(l,\rho,\phi, z)|^{2} = 0$ to determine where the particle density goes to zero. The two real solutions for the radii are given by
\begin{equation}
R_{\pm}^{2} =\frac{ \sigma M \pm \sqrt{\sigma^{2} M^{2}+4 \lambda g \mu}}{\lambda g}
\end{equation}
where $M= m \omega_{\perp}^{2}$ and $g=m^{2} \omega_{\perp}^{3}/\hbar ^{2}$.
A typical spatial form of the ground state BEC cloud in the mexican hat potential is shown in Fig.~\ref{Fig:WF} along one transverse direction ($\hat{x}$ or $\hat{y}$). The toroid shaped particle distribution should be clear from the figure.

\begin{figure}[ht]
\includegraphics[scale=1.0]{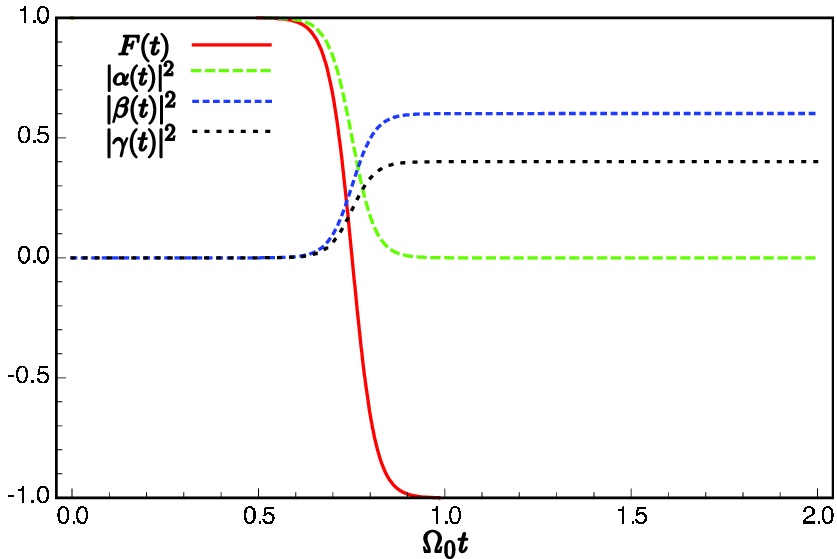}
\caption{\label{Fig:MexHatTransfer} Generation of the vortex state superposition in a Mexican hat trap: Results of the numerical solutions of the equations for STIRAP scheme shows the superposition 60:40 of the $\ket{+}$ and $\ket{-}$ vortex states. The Rabi frequency $\Omega_{0}=1$kHz and $\Delta = 100\Omega_{0}$.}
\end{figure}

The spatial part of the BEC wavefunction for different components is taken to be
\begin{align}
\psi_g (\mathbf{r}) &= \psi(0,\rho,\phi, z)\,, \nonumber \\
\psi_{v \pm} (\mathbf{r}) &= \psi(\pm \ell,\rho,\phi, z) \,.
\end{align}
Using the above one can evaluate the spatial integrals appearing in the Eq.~\eqref{Eq:RateEqBasicFormal} governing the population dynamics. We, furthermore, ignore kinetic energy of the system as we are working in the TF approximation, in this case and choose the optical field frequencies so that there is two-photon resonance, i.e. the two-photon detuning $\delta = 0$. The light fields $\Omega_{c}$ and $\Omega_{\pm}$ have the same saptio-temporal form as discussed in sec. \ref{sec:STIRAP}. Numerically solving the evolution equations, we get the transfer curves for the mexican hat potential depicted in Fig. \ref{Fig:MexHatTransfer}. We observe that the time-scale of transfer is about the same as in the case of harmonic trap. The figure depicts generation of a 60:40 superposition of $\ket{+2}$ and $\ket{-2}$ vortex states. Furthermore, it is important to note that any arbitrary superposition could be obtained in the mexican-hat trap; the generated vortex state would be naturally stable and would not disintegrate into single charge vortices.

\forget{From the magnitude of the spatial integrals we can conclude that the change in population of the levels over time does not affect the transfer significantly as the atom-light interaction has a much smaller timescale ($\mu$s) than the atom-atom interaction timescale (ms).}

\section{\label{Sec:Detection}Detection of the BEC vortex superposition}
\begin{figure}[ht]
\includegraphics[width=0.9\columnwidth]{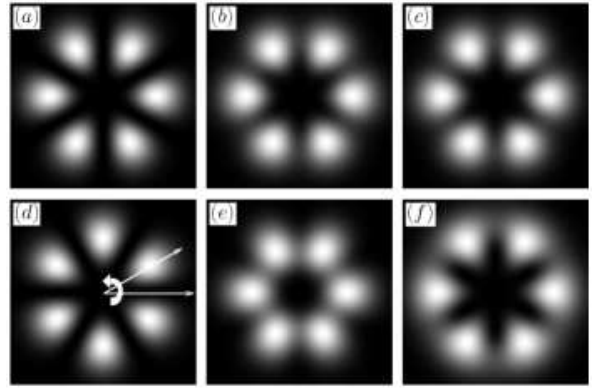}
\caption{\label{Fig:Interference} Detection of the vortex-superposition through characteristic interference of a general normalized state $\alpha \ket{+\ell} + \beta\, \re^{\ri \theta}\ket{-\ell}$. $\ell=3$ and $\theta=0$ unless specified otherwise. $(a)$ 
$\alpha^2:\beta^2=1:1\,;V=1$, giving $m=2\ell$ lobes in the interference pattern. 
$(b)$ $\alpha^2:\beta^2=0.1:0.9\,; V=0.6$.  $(c)$ $\alpha^2:\beta^2=0.9:0.1\,; V=0.6$. 
Patterns $(b)$ and $(c)$ are identical. $(d)$ Phase determination: $\alpha^2:\beta^2=1:1\,,\theta=\pi; V=1$, notice rotation of the pattern with respect to that of $(a)$ by an angle $\theta/m$. After adding one more unit of OAM in the superposition to arrive at $\alpha \ket{+\ell+1} + \beta\, \re^{\ri \theta}\ket{-\ell+1}$ distinct patterns $(e)$ and $(f)$ are obtained instead of $(b)$ and $(c)$.}
\end{figure}

The resulting vortex-superposition state could be detected by imaging its particle density distribution, which is proportional to an interference  of its components. Fig.~\ref{Fig:Interference} shows the $x$-$y$ cross section of this interference pattern for a particular vortex state $\ket{\Psi}_v=\alpha \ket{\ell=+3} + \beta\, \re^{\ri \theta} \ket{-\ell=-3}$. 
The spatial profile of the interference pattern can be obtained by evaluating $|\braket{\mathbf{r}}{\Psi_v}|^2=A[1 + 2 \alpha \beta \cos(2 \ell \phi -\theta)]$ where $A = |\Psi_{v\pm}(r,\phi)|^2$ is just the toroid-shaped particle density distribution of the vortex states. The ``cosine''  term dictates the modulation of the particle density distribution along the azimuth with $2\ell$ oscillations as the azimuthal coordinate $\phi$ changes from $0$ to $2\pi$. The relative phase of the two components, $\theta$,  just causes offset( or rotation) of the interference pattern by an angle $\theta/(2\ell)$. Note that ${\mathbf r}$ is a two-dimensional vector signifying the position of a point in polar coordinate system $\{r,\phi\}$. The visibility of such a pattern can be readily arrived at via:
\begin{equation}
V = \frac{I_{max}-I_{min}}{I_{max}+I_{min}} = 2 \alpha \beta\,
\end{equation}
as $I_{max} = A(1 + 2 \alpha \beta)$ and $I_{min} =A( 1 - 2 \alpha \beta)$ are the extremal intensities.

In general, for such a superposition the spatial profile of the interference pattern contains $m= 2 \ell$ lobes. The visibility $V=2 \alpha \beta$, as determined above, gives a measure of the asymmetry in the amplitudes. The pair $\{\alpha, \beta\}$ can be determined using measured $V$ and the normalization condition $\alpha^2 + \beta^2 =1$. However, this does not assign the amplitudes to the states $\ket{+}$ or $\ket{-}$ with certainty, as the patterns in $(b)$ and $(c)$ are identical.  We propose shining a OAM $\ell=+1$ light of $\sigma_+$ polarization to obtain the vortex state  $\alpha \ket{3+1} + \beta \re^{\ri \theta} \ket{-3+1}$, the resulting interference pattern is shown in Fig.~\ref{Fig:Interference} $(e)$ and $(f)$, which now clearly differentiates between the two amplitude values that gave same visibility in $(b)$ and $(c)$. The phase difference $\theta$ causes rotation of the whole pattern by an amount $\theta/m$  as shown in $(d)$ of the figure.  Existing  schemes for detecting vortex states~\cite{Bolda:1998a} could also be extended to detect a superposition of vortex states. 

\section{\label{Sec:Conclusion}Conclusion}
To summarize, we have studied two different mechanisms for transfer of superposition of optical angular momentum of light to vortices in BEC.  An interferometric scheme for generation of an arbitrary superposition of two different OAM states of light is also presented in great detail. We also discussed a couple of trap configurations and showed that the OAM transfer from light to BEC works independently of the trapping potential and despite inter-atomic interactions present in the BEC. The applications of the techniques discussed here are to a memory for the OAM states of light with further applications in the quantum or even classical communications protocols using OAM states of light. The superposition states of vortices in BEC could also be used as a qubit for quantum information processing; however, further work is clearly necessary on that front. Further, various advantages of using atoms versus photons for interferometric metrology are discussed in Ref.~\cite{Dowling:1998}.  The superposition of counter-rotating currents in the BEC could also be used for inertial sensing, especially as a gyroscope. Several advantages offered by a gyroscope based on superpositions of counter-rotating vortex structures would be the tunability of the effective de-Broglie wavelength by choosing appropriate atomic masses and the angular velocity and phase sensitivity via the choice of the quantized angular momentum of the atoms.

\section{Acknowledgments}
We would like to acknowledge support from the Disruptive Technologies Office, and the Army Research Office.

\appendix
\section{\label{App:MachZehnder}Detailed Description of the Mach Zender Interferometer}
In this appendix we offer mathematical details of the interferometric scheme for generation of the OAM superposition for light. We suggest references \cite{Zeilinger:1981,Jaroszkiewicz:2004, Holbrow:2002} to an interested reader for further details on treatment of the optical elements.

\begin{figure}[ht]
\includegraphics[scale=0.5]{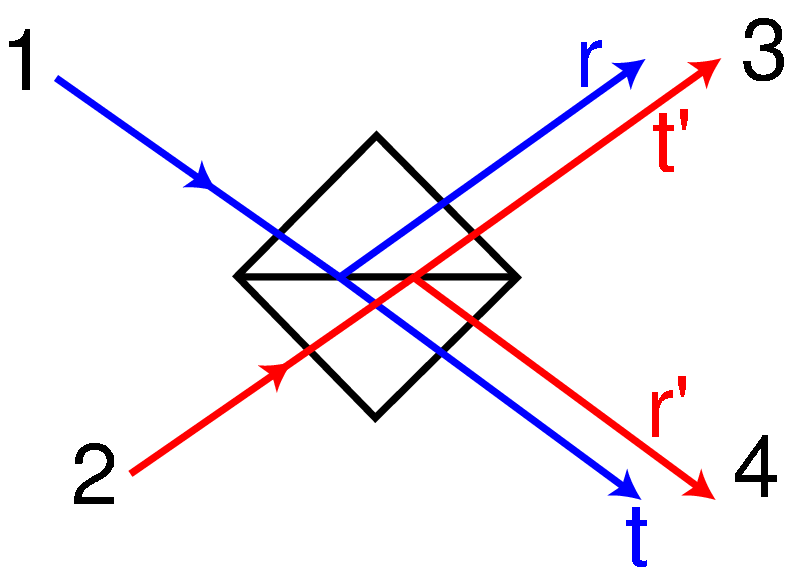}
\caption{\label{Fig:BS} A schematic representation of an ideal beam splitter with input ports 1 and 2 and output ports 3 and 4. The amplitudes for reflection and transmission from port 1 are, respectively $r$ and $t$. The corresponding amplitudes for a photon entering port 2 are $r'$ and $t'$.}
\end{figure}
We chose a basis state representation as
\begin{align}
\ket{1} \equiv \(\begin{matrix} 1 \\ 0 \end{matrix}\), 
\quad \ket{2} \equiv \(\begin{matrix} 0 \\ 1 \end{matrix}\)\,;
\nonumber \\
\ket{3} \equiv \(\begin{matrix} 1 \\ 0 \end{matrix}\), 
\quad \ket{4} \equiv \(\begin{matrix} 0 \\ 1 \end{matrix}\)\,.
\end{align}
In this representation the general beam 
splitter matrix can be written as
\begin{equation}
R=\(\begin{matrix}
r & t' \\
t & r' \\
\end{matrix}\)\,.
\end{equation}
Were $r$ and $t$ are the reflection and transmission amplitudes for the input at port 1, and $r'$ and $t'$ are the parameters for the port 2 (See Fig.~\ref{Fig:BS}). 
Noting that the matrix $R$ should be unitary, meaning, $R^\dagger=R^{-1}$ we obtain
\begin{equation}
\(\begin{matrix}
r^* & t^* \\ 
t'^* & r'^*
\end{matrix}\) = \frac{1}{rr' - tt'}
\(\begin{matrix}
r' & -t' \\ 
-t & r
\end{matrix}\)
\end{equation}
The determinant of a unitary matrix has a modulous of one, therefore we have $rr'-tt'={\rm e}^{\ri \gamma}$. However, this factor multiplies all the elements of the matrix, thus we can safely choose it to be 1,  given by the choice of $\gamma=0$. By equating the corresponding elements on the RHS and LHS of the above equation we obtain
\begin{equation}
r' = r^* \text{  and } t' = - t^*
\label{Eq:rtamp}
\end{equation}
If we rewrite these factors as complex exponentials, $|r| {\rm e }^{\ri \delta_r}$,   $|t| {\rm e }^{\ri \delta_t}$, $|r'| {\rm e }^{\ri \delta_{r'}}$ and $|t'| {\rm e }^{\ri \delta_{t'}}$. by division of the above equalities we obtain
\begin{equation}
\frac{|t|}{|r|}  {\rm e}^{\ri (\delta_{t} - \delta_{r})} = - \frac{|t'|}{|r'|}  {\rm e}^{-\ri(\delta_{t'} - \delta_{r'})}\,.
\end{equation}
From \eqref{Eq:rtamp} we can see that $|t|=|t'|$ and $|r|=|r'|$; using this in the above equation we obtain,
\begin{equation}
\delta_{t} - \delta_{r} + \delta_{t'} - \delta_{r'} = \pi\,.
\end{equation}
For the case of a symmetric beam splitter, which has the same effect on a beam incident through port labeled 1 as on a beam incident through port labeled 2, $r=r'$ and $t=t'$, and we have $\delta_{t} - \delta_{r} + \delta_{t'} - \delta_{r'} = \pi/2$. Thus the transmitted wave leads the reflected wave in phase by $\pi/2$ radians. This is a general property of symmetric beam splitter~\cite{Zeilinger:1981}. This also implies that  $r$  and $r'$ are purely real and equal  and $t$ and $t'$ are purely imaginary and equal to each other. Let $r=r'=\tilde{r}$ and $t=t' = \ri\, \tilde{t}$ be the 
Thus the beam splitter matrix becomes
\begin{equation}
\label{Eq:RSym}
\tilde{R}=\(\begin{matrix}
\tilde{r} & \ri\,\tilde{t} \\
\ri\,\tilde{t} & \tilde{r} \\
\end{matrix}\)\,.	
\end{equation}
For the case of a 50-50 beam splitter, for which  $r = r'$ and $t=t'$ and also $|r|=|r'| = |t|=|t'|$, we have
\begin{equation}
\label{Eq:R5050}
\tilde{R}^{(50-50)}=\frac{1}{\sqrt{2}}\(\begin{matrix}
1 & \ri \\
\ri & 1  \\
\end{matrix}\)\,.	
\end{equation}
Now we consider the complete transformations for the Interferometer considered in Fig.~\ref{Fig:OAMS};
\begin{widetext}
\begin{align}
\(\begin{matrix} u_1 \\ u_2 \end{matrix}\) &= 
\(\begin{matrix}
\tilde{r} & \ri\,\tilde{t} \\
\ri\,\tilde{t} & \tilde{r} \\
\end{matrix}\)
\(\begin{matrix}
\ket{-\ell}\bra{\ell} & 0 \\
0 & 1 \\
\end{matrix}\)
\(\begin{matrix}
1 & 0 \\
0 & {\rm e}^{\ri \phi} \\
\end{matrix}\)
\(\begin{matrix}
\tilde{r} & \ri\tilde{t} \\
\ri\tilde{t} & \tilde{r} \\
\end{matrix}\)\(\begin{matrix} u_0 \ket{\ell} \\ 0 \end{matrix}\) 
\nonumber \\
&=
\(\begin{matrix}
\tilde{r}^2 \ket{-\ell}\bra{\ell} - \tilde{t}^2 {\rm e}^{\ri \phi}  & \ri\tilde{r}\tilde{t} \ket{-\ell}\bra{\ell}+ \ri\tilde{r}\tilde{t}{\rm e}^{\ri \phi} \\
\ri\tilde{r}\tilde{t}\ket{-\ell}\bra{\ell} + \ri\tilde{r}\tilde{t}{\rm e}^{\ri \phi} & -\tilde{t}^2\ket{-\ell}\bra{\ell} + \tilde{r}^2 {\rm e}^{\ri \phi} \\
\end{matrix}\)
\(\begin{matrix} u_0 \ket{\ell} \\ 0 \end{matrix}\)= \(\begin{matrix}  \tilde{r}^2 u_0 \ket{-\ell}  - \tilde{t}^2 {\rm e}^{\ri \phi} u_0\ket{\ell} \\   \ri\, \tilde{t} \tilde{r} \, u_0 \ket{-\ell}  + \ri \tilde{r}\tilde{t}  {\rm e}^{\ri \phi}\, u_0\ket{\ell} \end{matrix}\)\,.
\end{align}
Observing the output state, it can be easily seen that a general superposition could not be generated by this method. Instead we need only one beam-splitter to be imbalanced, meaning $|r| \neq |t|$, and the other one to be 50:50. However, we can still choose both the beam splitters to be symmetric, i.e., $r = r'$ and $t=t'$. Thus using  Eq.~\eqref{Eq:RSym} and Eq.~\eqref{Eq:R5050} we can attain the output state.
\begin{align}
\(\begin{matrix} u_1 \\ u_2 \end{matrix}\) &= 
\frac{1}{\sqrt{2}}\(\begin{matrix}
1 & \ri \\
\ri & 1 \\
\end{matrix}\)
\(\begin{matrix}
\ket{-\ell}\bra{\ell} & 0 \\
0 & 1 \\
\end{matrix}\)
\(\begin{matrix}
1 & 0 \\
0 & {\rm e}^{\ri \phi} \\
\end{matrix}\)
\(\begin{matrix}
\tilde{r} & \ri\tilde{t} \\
\ri\tilde{t} & \tilde{r} \\
\end{matrix}\)\(\begin{matrix} u_0 \ket{\ell} \\ 0 \end{matrix}\) 
\nonumber \\
&=\frac{1}{\sqrt{2}}
\(
\begin{matrix}
\tilde{r} \ket{-\ell}\bra{\ell} - \re^{\ri \phi}\, \tilde{t} & \ri \tilde{t} \ket{-\ell}\bra{\ell} + \ri \tilde{r}\,\re^{\ri \phi} \\
\ri \tilde{r} \ket{-\ell}\bra{\ell} + \ri \tilde{t}\,\re^{\ri \phi} & -\tilde{t}\ket{-\ell}\bra{\ell} + \tilde{r} \re^{\ri \phi}
\end{matrix}
\)\(\begin{matrix}
u_0 \ket{\ell} \\ 0
\end{matrix}
\) = 
\frac{1}{\sqrt{2}}u_0
\(
\begin{matrix}
 \tilde{r} \ket{-\ell} - \re^{\ri \phi}\, \tilde{t}\ket{\ell}  \\
 \ri \tilde{r} \ket{-\ell} + \ri \tilde{t} \re^{\ri \phi}\ket{\ell}
\end{matrix}
\)\,.
\end{align}
Thus, by choosing $\tilde{t}=a_{+}$ and $\tilde{r}=a_{-}$, with the condition that $\tilde{r}^2 + \tilde{t}^2=1$, and $\re^{\ri \phi}=-1$ we obtain the state
$u_0( a_{+} \ket{\ell} + a_{-} \ket{-\ell})/\sqrt{2} $, which is a general superposition state.
\end{widetext}

\forget{\section{Derivation of the rate equations for Optical Coupling of multi-component BEC}

\section{Adiabatic Elimination}}

\section{The spatial Integrals}
The various coordinate integrals needed to eliminate the spatial parts of the BEC spinors to arrive at the population evolution equations are given below. The definitions given below are independent of the trapping potential or the ansatz used for the spatial profile of the wavefunction. 
\begin{align}
T_{\rm g}&=\frac{1}{\hbar}\int \psi_{\rm g}^*({\mathbf r})\, \mathcal{T}\, \psi_{\rm g}({\mathbf r})\,\,  \rd^3 {\mathbf r}\,, \nonumber \\
V_{\rm g}&=\frac{1}{\hbar}\int \psi_{\rm g}^*({\mathbf r})\, \mathcal{V}\, \psi_{\rm g}({\mathbf r})\,\,  \rd^3 {\mathbf r}\,, \nonumber \\
T_{\pm}(\ell)&=\frac{1}{\hbar}\int \psi_{{\mathrm v}\pm}^*(\pm\ell,{\mathbf r})\,  \mathcal{T}\, \psi_{\rm v\pm}(\pm\ell,{\mathbf r})\,\,  \rd^3 {\mathbf r}\,, \nonumber \\
V_{\pm}(\ell)&=\frac{1}{\hbar}\int \psi_{\rm v\pm}^*(\pm\ell,{\mathbf r})\,  \mathcal{V}\, \psi_{\rm v\pm}(\pm\ell,{\mathbf r})\,\,  \rd^3 {\mathbf r}\,, \nonumber \\  
I_{\rm gg} &=\frac{\eta}{\hbar}\int |\psi_{\rm g}({\mathbf r})|^2|\psi_{\rm g}({\mathbf r})|^2\,\,  \rd^3 {\mathbf r}\,, \nonumber \\
I_{\rm g\pm}(\ell)&=\frac{\eta}{\hbar}\int |\psi_{\rm v\pm}(\ell,{\mathbf r})|^2|\psi_{\rm g}({\mathbf r})|^2\,\,  \rd^3 {\mathbf r}\,, \nonumber \\  
I_{++}(\ell)&=\frac{\eta}{\hbar}\int |\psi_{\rm v+}(\ell,{\mathbf r})|^2|\psi_{\rm v+}(\ell,{\mathbf r})|^2\,\,  \rd^3 {\mathbf r}\,, \nonumber \\
I_{--}(\ell)&=\frac{\eta}{\hbar}\int |\psi_{\rm v-}(\ell,{\mathbf r})|^2|\psi_{\rm v-}(\ell,{\mathbf r})|^2\,\,  \rd^3 {\mathbf r}\,, \nonumber \\
I^{(2\ell)}_{\rm gg}(\ell)&=\int \psi_{\rm g}^*({\mathbf r})\,\(\frac{\sqrt{2} r}{w}\)^{2|\ell|}\,\psi_{\rm g}({\mathbf r})\,\,  \rd^3 {\mathbf r} \,, \nonumber \\
I^{(\ell)}_{\rm g\pm}(\ell)&=\int \psi_{\rm g}^*({\mathbf r})\,\re^{\mp\ri \ell \phi} \(\frac{\sqrt{2} r}{w}\)^{|\ell|} \psi_{\rm v\pm}(\ell,{\mathbf r})\,\,  \rd^3 {\mathbf r}\,, \nonumber \\
I^{(\ell)}_{\rm \pm g}(\ell)&=\int \psi_{\rm v\pm}^*(\ell,{\mathbf r})\,\re^{\pm\ri \ell \phi} \(\frac{\sqrt{2} r}{w}\)^{|\ell|} \psi_{\rm g}({\mathbf r})\,\,  \rd^3 {\mathbf r}\,.
\end{align}
The values of the above spatial integrals, for specific vortex states corresponding to the charge of $\ell=\pm2$ and the harmonic trapping potential, are given below. 
\begin{align}
T_{\rm g}^{(H)}  &= \frac{\hbar}{4m}\,\( \frac{1}{L_z^{2}} + \frac{2}{{L_{\perp}}^2} \right) = \frac{1}{4} \omega_z  + \frac{1}{2}  \omega_{\perp}=V_{\rm g}^{(H)} \,,\nonumber \\
T_{\pm}^{(H)}(2) &= \frac{\hbar}{4m}\,\( \frac{1}{L_z^{2}} + \frac{6}{{L_{\perp}}^2}
      \right)= \frac{1}{4} \omega_z  + \frac{3}{2}  \omega_{\perp}=V_{\pm}^{(H)} (2)\,, \nonumber \\      
I_{\rm gg}^{(H)} &= \frac{\eta}{{(2\pi) }^{\frac{3}{2}}\,{\hbar~L_z}\,
    {{L_{\perp}}}^2} = 4 \kappa\,,  \nonumber \\
I_{\rm g+}^{(H)}(2) &= \frac{\eta}{4{(2\pi) }^{\frac{3}{2}}\,{\hbar~L_z}\,
    {{L_{\perp}}}^2}= {\kappa}=I^{(
    H)}_{\rm g-}(2)\,, \nonumber \\
I_{++}^{(H)}(2)&= \frac{3\eta}{8{(2\pi) }^{\frac{3}{2}}\,{\hbar~L_z}\,{{L_{\perp}}}^2}= \frac{3}{2}\kappa= I^{(H)}_{--}(2)= I^{(H)}_{+-}(2)\,, \nonumber \\
I^{(4)(H)}_{\rm gg}&= 2 L_{\perp}^4 =2 \({\hbar}/{m \omega_{\perp}}\)^2 \nonumber \\
I^{(2)(H)}_{\rm g\pm}(\ell)&= \sqrt{2} L_{\perp}^2 =I^{(2)(H)}_{\rm \pm g}(\ell). 
\end{align}
Note also that at this stage we have used various properties of the quantum harmonic oscillator as summarized below:
\begin{align}
\frac{\hbar^2}{2 m L_{\perp}^2} = \frac{1}{2} m \omega_{\perp}^2, \text{ i.e., } L_{\perp} = \sqrt{\frac{\hbar}{m \omega_{\perp}}}, \text{ and }L_{z} = \sqrt{\frac{\hbar}{m \omega_{z}}},\nonumber \\  \nonumber 
\end{align}
The corresponding integrals for a the mexican hat trap are evaluated numerically for the parameter values $\sigma = 2.0$ and $\lambda = 0.005$. Also, as we deal with the Thomas-Fermi ground states for the mexican hat trap the kinetic energy is ignored.

This completes the description of the spatial integrals, which  allow us to obtain  the time-dependent population equations~\eqref{Eq:RateEqBasicFormal}. Those could be solved as discussed in the main text of the paper to study the transfer techniques for the vortex superpositions from the light to the atom.


\end{document}